\documentclass[preprint,aps,prl,onecoloumn,superscriptaddress]{revtex4}
\usepackage{graphicx}
\usepackage{times}
\usepackage{amsmath}
\usepackage{textcomp}

\begin{document}

\title{Josephson supercurrent through a topological insulator surface state}

\author{M. Veldhorst}
\affiliation{Faculty of Science and Technology and MESA+ Institute for Nanotechnology, University of Twente, 7500 AE Enschede, The Netherlands}
\author{M. Snelder}
\affiliation{Faculty of Science and Technology and MESA+ Institute for Nanotechnology, University of Twente, 7500 AE Enschede, The Netherlands}
\author{M. Hoek}
\affiliation{Faculty of Science and Technology and MESA+ Institute for Nanotechnology, University of Twente, 7500 AE Enschede, The Netherlands}
\author{T. Gang}
\affiliation{Faculty of Science and Technology and MESA+ Institute for Nanotechnology, University of Twente, 7500 AE Enschede,
The Netherlands}
\author{X.L. Wang}
\affiliation{Institute for Superconducting and Electronic Materials, University of Wollongong, Wollongong, NSW, 2522, Australia}
\author{V.K. Guduru}
\affiliation{High Field Magnet Laboratory, Institute for Molecules and Materials, Radboud University Nijmegen, 6525 ED Nijmegen, The Netherlands}
\author{U. Zeitler}
\affiliation{High Field Magnet Laboratory, Institute for Molecules and Materials, Radboud University Nijmegen, 6525 ED Nijmegen, The Netherlands}
\author{W.G. v.d. Wiel}
\affiliation{Faculty of Science and Technology and MESA+ Institute for Nanotechnology, University of Twente, 7500 AE Enschede,
The Netherlands}
\author{A.A. Golubov}
\affiliation{Faculty of Science and Technology and MESA+ Institute for Nanotechnology, University of Twente, 7500 AE Enschede, The Netherlands}
\author{H. Hilgenkamp}
\altaffiliation[Also at ]{Leiden Institute of Physics, Leiden University, P.O. Box 9506, 2300 RA Leiden, The Netherlands}
\affiliation{Faculty of Science and Technology and MESA+ Institute for Nanotechnology, University of Twente, 7500 AE Enschede, The Netherlands}
\author{A. Brinkman}
\affiliation{Faculty of Science and Technology and MESA+ Institute for Nanotechnology, University of Twente, 7500 AE Enschede, The Netherlands}\date{\today}

\begin{abstract}
\textbf{Topological insulators \cite{Zhang2006, Konig2007, Fu2007, Zhang2009N, Qi2009, Hsieh2008, Chen2009, Hsieh2009N, Peng2009, Cheng2010, Zhang2009P} are characterized by an insulating bulk with a finite band gap and conducting edge or surface states, where charge carriers are protected against backscattering. These states give rise to the quantum spin Hall effect \cite{Konig2007} without an external magnetic field, where electrons with opposite spins have opposite momentum at a given edge.
The surface energy spectrum of a three-dimensional topological insulator \cite{Fu2007, Zhang2009N} is made up by an odd number of Dirac cones with the spin locked to the momentum. The long-sought yet elusive Majorana fermion \cite{Majorana1937} is predicted to arise from a combination of a superconductor and a topological insulator \cite{Fu2008, Nilsson2008, Tanaka2009}. An essential step in the hunt for this emergent particle is the unequivocal observation of supercurrent in a topological phase. Here, we present the first measurement of a Josephson supercurrent through a topological insulator. Direct evidence for Josephson supercurrents in superconductor (Nb) - topological insulator (Bi$_2$Te$_3$) - superconductor e-beam fabricated junctions is provided by the observation  of clear Shapiro steps under microwave irradiation, and a Fraunhofer-type dependence of the critical current on magnetic field. The dependence of the critical current on temperature and length shows that the junctions are in the ballistic limit. Shubnikov-de Haas oscillations in magnetic fields up to 30~T reveal a topologically non-trivial two-dimensional surface state. We argue that the ballistic Josephson current is hosted by this surface state despite the fact that the normal state transport is dominated by diffusive bulk conductivity. The lateral Nb-Bi$_2$Te$_3$-Nb junctions hence provide prospects for the realization of devices supporting Majorana fermions \cite{Hasan2010}.}
\end{abstract}

\maketitle

\begin{figure}[b]
	\centering 
		\includegraphics[width=0.4\textwidth]{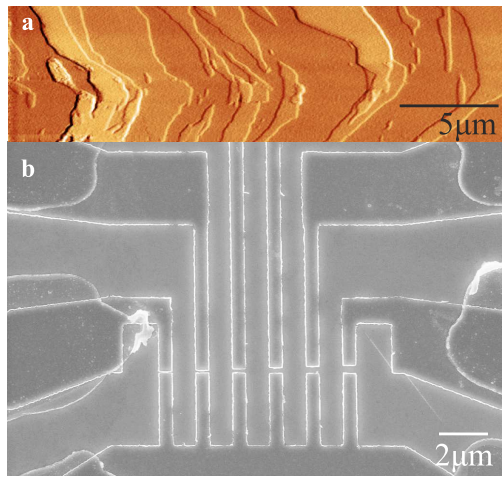}
		\caption{\textbf{ E-beam lithographically defined Nb electrodes on exfoliated Bi$_2$Te$_3$. (a)} Atomic force microscopy image of an exfoliated Bi$_2$Te$_3$ surface. The step edges are 1.0~nm high, corresponding to the Bi$_2$Te$_3$ quintuple unit cell. These nanometer flat surfaces span an area up to 50 $\times$ 50~\textmu m$^2$. \textbf{(b)} Scanning electron microscopy image of Nb-Bi$_2$Te$_3$-Nb Josephson junctions. The Nb superconductor strips are defined by e-beam lithography on a 200 nm thick exfoliated Bi$_2$Te$_3$ flake and connected to large Nb electrodes fabricated by photolithography. The junctions have a width of 500~nm and a length of (from left to right) 50, 100, 150, 200, 250 and 300~nm. A supercurrent has been identified at 1.6~K in all junctions up to 250~nm.}  
		\label{fig:1}
		\vspace{-15pt} 
\end{figure}

\begin{widetext}
\begin{figure}[b]
	\centering 
		\includegraphics[width=1\textwidth]{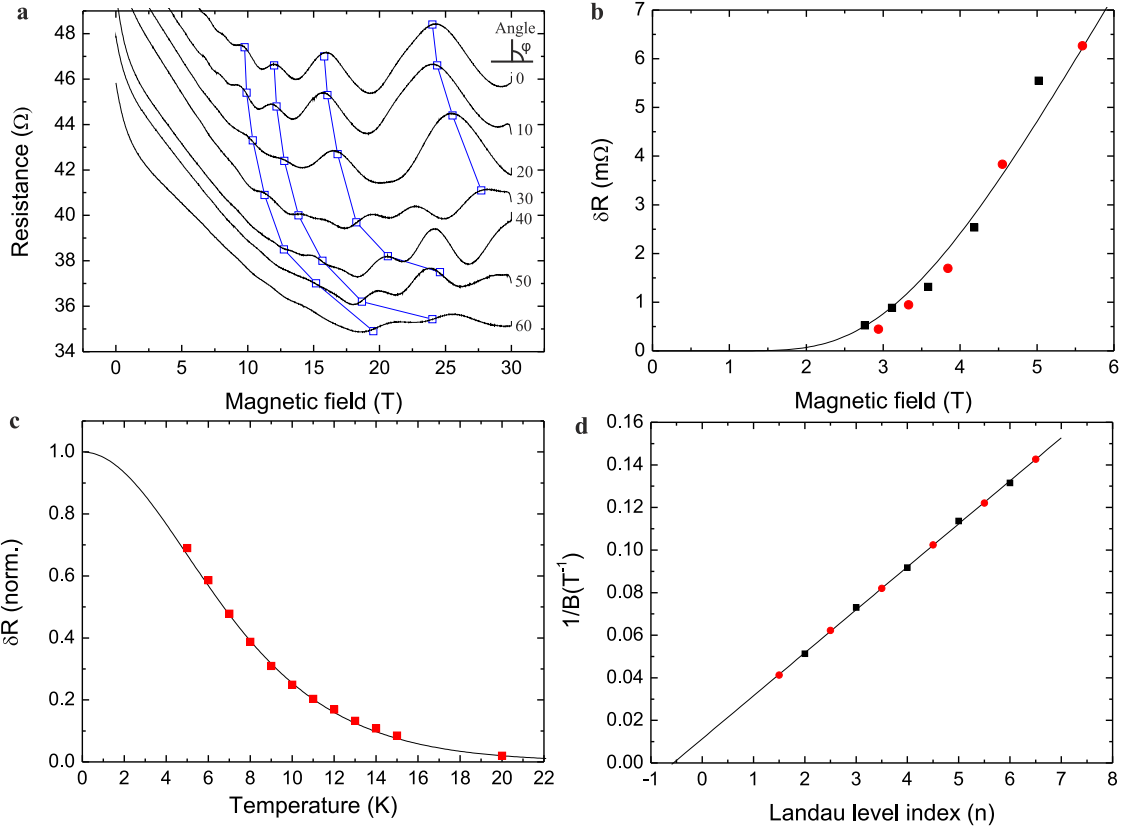}
		\caption{\textbf{Magnetoresistance oscillations of the Bi$_2$Te$_3$ surface states. (a)} Angle dependence of the Shubnikov-de Haas oscillations, at 4.2~K, (top to bottom: the angle between the magnetic field and the surface normal $\phi$ = $0^\circ$ to $60^\circ$ in steps of $10^\circ$) after linear background subtraction. The square markers represent the expected shift of the maxima in magnetic field for a two dimensional system, given by $B_\bot=B\cos(\phi)$, indicating that the oscillations arise from surface states. \textbf{(b)} Oscillation amplitude dependence on the magnetic field (minima as black squares and maxima as red circles); linear backgrounds have been subtracted leaving the oscillating part. We deduce a Dingle temperature $T_D=$ 1.65~K from fitting the increase in oscillation amplitude with increasing field. Only the magnetoresistance in small fields is considered, where the peak splitting is negligible. \textbf{(c)} Oscillation amplitude dependence on temperature; linear backgrounds have been subtracted. The effective mass $m^*=0.16~m_0$ is estimated by fitting the decrease of the magnetoresistance oscillations as function of temperature (peaks at least up to 9~T in both the conductance and resistance result in the same effective mass). \textbf{(d)} The $1/B$ values versus the Landau level index $n$ intersect at $n=-0.5$, consistent with a half filled lowest Landau level as expected for a Dirac cone.}  
		\label{fig:2}
		\vspace{-15pt} 
\end{figure}
\end{widetext}

\begin{figure}[]
	\centering 
		\includegraphics[width=1\textwidth]{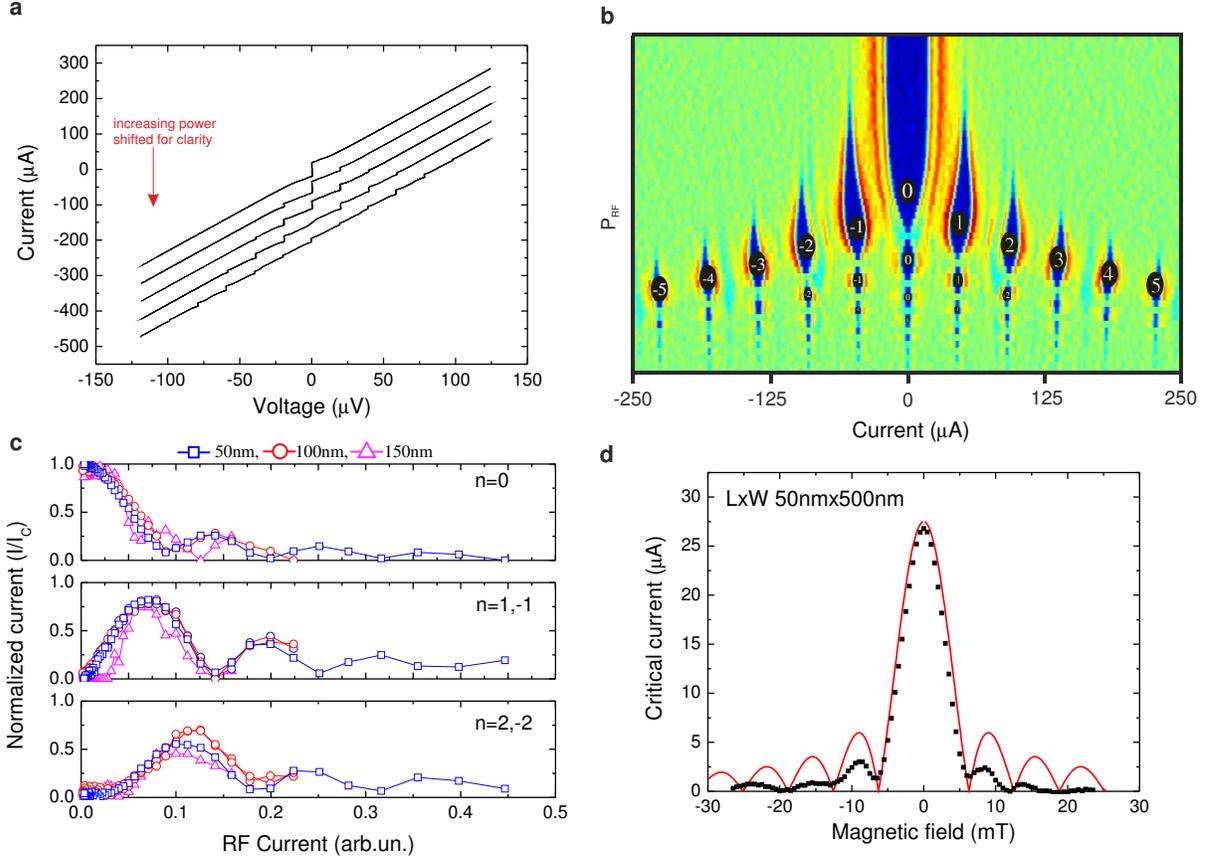}
		\caption{\textbf{Josephson effects. (a)} Current voltage characteristics of a Nb-Bi$_2$Te$_3$-Nb Josephson junction (length l=50~nm) upon increased 10.0~Ghz RF irradiation power (curves are shifted for clarity). The temperature is 1.6~K. Clear Shapiro steps are observed at multiples of $V=\frac{h}{2e}f_{\textrm{RF}}=20.7~$\textmu V. \textbf{(b)} Differential resistance, $dV/dI$, plotted on a color scale as a function of the bias current, $I$, and the microwave excitation power, $P_{RF}$. The numbers correspond to the n-th order Shapiro step. \textbf{(c)} Power dependence of the first three Shapiro steps. Shapiro steps are resolved in the junctions with length up to 150~nm. \textbf{(d)} Critical current dependence on the magnetic field fitted with $I_C=I_0 \left|  \text{sinc} \left(  \frac{\pi \Phi}{\Phi_0} \right) \right|$, $I_0$ the critical current at zero field, $\Phi$ the flux and $\Phi_0$ the flux quantum. The temperature is 260~mK.}
		\label{fig:3}
		\vspace{-15pt} 
\end{figure}

\begin{figure}[]
	\centering 
		\includegraphics[width=0.8\textwidth]{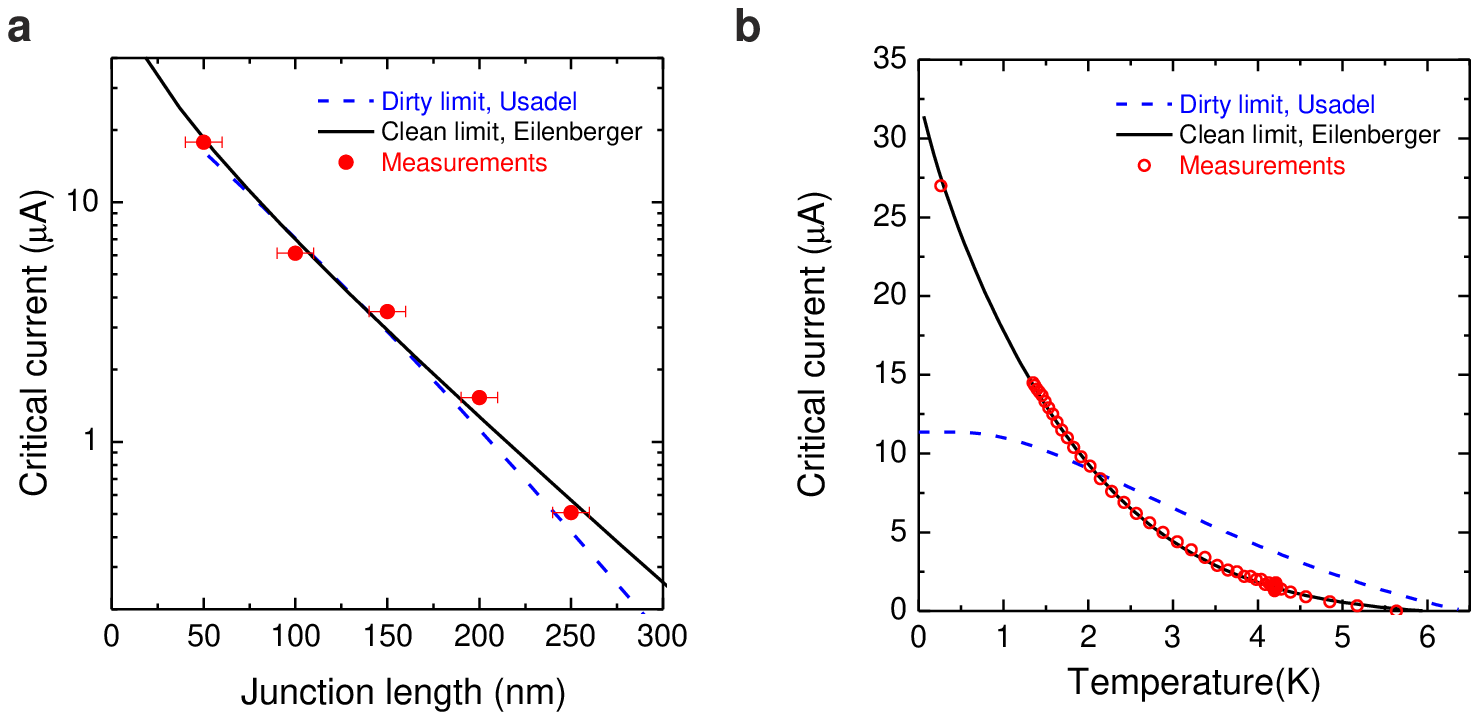}
		\caption{\textbf{Temperature and length dependence of the critical current; demonstration of the ballistic nature of the junctions. (a)} Scaling of the junction critical current with the electrode separation length at 1.6~K. \textbf{(b)} Temperature dependence of the critical current of the 50~nm junction. The measured data (red squares) can only be consistently fitted in both cases with Eilenberger theory for ballistic junctions (dashed black curves), while Usadel theory for diffusive junctions (dotted blue curves) cannot provide a good fit to the temperature dependence of the critical current. See the Supplementary Information for details. The obtained clean limit coherence length $\xi=75~$nm at 1.6~K. }  
		\label{fig:4}
		\vspace{-15pt} 
\end{figure}

Almost simultaneous to the theoretical prediction of 3D topological insulator states in the bismuth compounds Bi$_{1-x}$Sb$_x$, Bi$_2$Se$_3$ and Bi$_2$Te$_3$ \cite{Zhang2006}, angle-resolved photoelectron spectroscopy indeed revealed a linear dispersion and a helical structure of the Dirac cone at the surface of these compounds  \cite{Hsieh2008, Chen2009, Hsieh2009N}. Soon after, the topological nature of the surface states was confirmed by transport studies such as Aharonov-Bohm oscillations in Bi$_2$Se$_3$ nanoribbons  \cite{Peng2009}, Shubnikov-de-Haas oscillations in Bi$_2$Te$_3$ \cite{Qu2010, Xiu2011, Analytis2010, Xiong2011, Taskin2011}, scanning tunneling spectroscopy of the square-root magnetic field dependence of the Landau level spacing \cite{Cheng2010} and interference effects resulting from impurity scattering \cite{Zhang2009P}. Now the existence of the topological surface states has been established it is time to study the interaction with other materials. Efforts have been made to contact a topological insulator to a superconductor in the search for the Majorana fermion. First attempts \cite{Dumin2011, Sacepe2011} indicated the presence of a supercurrent through a topological insulator, although no Josephson effects have been observed. In this Letter we  demonstrate that it is possible to induce a Josephson supercurrent in the surface state of the topological insulator Bi$_2$Te$_3$, as evidenced by the two hallmarks of the Josephson effect: a Fraunhofer-like magnetic field modulation of the critical current and the appearance of Shapiro steps upon microwave irradiation, resulting from the dc and ac Josephson effects, respectively.

We have fabricated polycrystalline Bi$_2$Te$_3$ samples with a common $c$-axis orientation using the Czochralski method as described elsewhere \cite{Li2010}. From these, flakes were produced by mechanical cleaving, proven to be a powerful tool for the fabrication of a few quintuple layer thin Bi$_2$Te$_3$ flakes \cite{Teweldebrhan2010}. Figure \ref{fig:1}a shows an atomic force microscopy image of the surface of typical exfoliated Bi$_2$Te$_3$ flakes. We have performed experiments on large flakes with atomically flat terraces, extending over several micrometers. The Bi$_2$Te$_3$ quintuple layer step edges are clearly visible as shown in Fig. \ref{fig:1}a. The exfoliated flakes have a thickness of 20~nm $-$ 2~$\hbox{\textmu}$m. Van der Waals forces bind the flakes to the Si-substrate, enabling subsequent  sputtering  of Nb electrodes by standard lift-off photolithography. E-beam lithography is used in a second step to define  smaller Nb nanostructures by lift-off sputter deposition on top of the flake, connected to the larger Nb electrodes fabricated with photolithography. A scanning electron microscopy image of such a device is given in Fig. \ref{fig:1}b. 

To characterize the electronic properties of our devices and to track them down to the surface states of a topological insulator we have performed magnetotransport experiments on Bi$_2$Te$_3$ flakes in a Van der Pauw geometry in fields up to 30~T. From the low field Hall coefficient $R_H=7.5\times10^{-8}~\Omega$m/T  and the zero-field resistivity  $\rho=300~$\textmu$ \Omega$cm we deduce a bulk conductivity channel with an n-type carrier concentration $n=8.3\times10^{19}~$cm$^{-3}$ and  mobility  $\mu$=250~cm$^2$/Vs corresponding to an electron mean free path $l_e=22~$nm for a parabolic band.  This low mobility would not allow the observation of quantum oscillations since the necessary condition, $\mu B \gg 1$, is far from being fulfilled in the magnetic fields applied. As shown in Fig. \ref{fig:2}a, we nevertheless observe clear Shubnikov-de Haas oscillations oscillations in the resistance implying another conduction channel with larger mobility. From the angle dependence of the position of the peaks we conclude that this additional channel is of two-dimensional nature.

The oscillatory contribution to the resistance of such a two-dimensional system can be written as  \cite{Lifshitz1956}
\begin{equation}\label{LK}
R_{xx} \propto \frac{\lambda}{\sinh{\lambda}} e^{-\lambda _D} \cos \left( \frac{2\pi E_F}{\hbar \omega_c} + \pi + \varphi_{B} \right)
\end{equation}
where, $\omega_c$ is the cyclotron frequency,
$E_F$ is the Fermi energy, $\lambda=2\pi^2k_BT/\hbar\omega_c$, $\lambda_D=2\pi^2k_BT_D/\hbar\omega_c$ with $T_D$ the Dingle temperature, and $\varphi_{B}$ is the Berry phase.
When plotting the $1/B$-positions of the minima and maxima of $R_{xx}$ (Fig. \ref{fig:2}d) as a function of the Landau level index it already becomes clear that the corresponding $1/B$-positions do not extrapolate to $\varphi_{B}=0$ as one would expect from an ordinary system, but is rather shifted by $\frac{1}{2}$ implying that $\varphi_{B}=\pi$, consistent with a half-filled zeroth Landau level that is present in a topological surface state \cite{Qu2010, Xiu2011, Analytis2010, Xiong2011, Taskin2011,Brune2011}, similar to graphene ~\cite{Novoselov2005, Zhang2005}. From the slope in Fig.~\ref{fig:2}d we infer the carrier concentration of this surface state of $n=1.2\times10^{12}~$cm$^{-2}$; an effective mass $m^*=0.16~m_0$ and a Dingle temperature of $T_D=1.65~$K (corresponding to a  mobility  $\mu=8300~$cm$^2$/Vs) follow from the temperature and field dependence of the Shubnikov-de Haas oscillations amplitudes (Figs.~\ref{fig:2}b and ~\ref{fig:2}c). This results in a $v_F=1.4\times10^5~$m/s with $l_e=105~$nm. 

In high magnetic fields two beating frequencies are present, as Xiong \textit{et al.} \cite{Xiong2011} also observed in their study on the topological surface states of Bi$_2$Te$_2$Se. A double frequency has also been observed in the 3D topological insulator strained HgTe \cite{Brune2011}, where it was concluded to originate from the top and bottom topological surface states, with different carrier densities due to the difference in electrostatic environment (substrate and vacuum). As can be seen in Fig. \ref{fig:2}a, the peak splitting becomes more pronounced with increasing parallel field, though it is already visible in the derivative of the magnetoresistance at smaller angles.

The superconducting Nb-Bi$_2$Te$_3$-Nb junctions, as depicted in Fig.~\ref{fig:1}b, show Shubnikov-de Haas oscillations at high magnetic fields with the same frequency, revealing the presence of topological surface states in these devices as well. The critical current of the Nb electrodes on top of the Bi$_2$Te$_3$ is larger than 30~mA for a strip of 500~nm width and 70~nm height. The junctions have a metallic temperature dependence, but upon cooling below 6.5~K, the resistance vanishes completely and a supercurrent is observed. Figure \ref{fig:3}a shows a current-voltage characteristic at 1.6~K showing a clear supercurrent of $I_c=18~\mu$A. To test whether this is a true Josephson supercurrent we performed phase-sensitive experiments. Figure \ref{fig:3}d shows the Fraunhofer pattern from the critical current dependence on the magnetic field due to the dc Josephson effect. The smaller sidelobes follow from the small ratio between the length and the width of the junction, as predicted by \cite{Barzykin1999}. Due to the ac Josephson effect, microwave irradiation should lead to Shapiro steps. Figure \ref{fig:3}a shows the $IV$-curve upon 10.0~GHz microwave irradiation. Clear steps are observed at integer values of $V=h/2ef_{\textrm{RF}}=20.7~$\textmu V. A colormap of the conductance dependence on the power and the current is shown in Fig. \ref{fig:3}b, visualizing the evolution of the Shapiro steps. In Fig. \ref{fig:3}c we have plotted the power dependence of the thirst three steps, following the expected Bessel function dependence. 

Now that we have confirmed  the Josephson nature of the devices, we can discuss in which band of the Bi$_2$Te$_3$ the proximity effect is induced. Figure \ref{fig:4} shows the temperature and length dependence of the critical current of the junctions. While the scaling with length can be described by diffusive as well as ballistic transport theory (see Suppl. Information for details), the temperature dependence of the junctions is clearly far from the diffusive limit and can only be fitted by the Eilenberger theory for ballistic junctions \cite{Brinkman2000, Galaktionov2002}. The bulk mean free path, $l_e=22~$nm, is too small to explain the ballistic nature of the supercurrent in the junctions. However, the surface conduction band has a larger mean free path $l_e=105$~nm. Noting that the mean free path obtained from Shubnikov-de Haas oscillations is usually an underestimate \cite{Dietl1978}, we conclude that the ballistic Josephson supercurrent is carried by the topological surface states of Bi$_2$Te$_3$. As we discuss in more detail in the Suppl. Information, the small $I_cR_n$ product (reduced to about 2$\%$) follows automaticaly from the large bulk shunt. The clean limit fit with $T_c = 6.5~$K gives a coherence length $\xi = \frac{\hbar v_F}{2 \pi k_B T}=75$~nm at 1.6~K, implying that $v_F=1.0\times10^5$~m/s.  This Fermi velocity is comparable to the value obtained from the magnetoresistance oscillations of the surface states ($1.4\times10^5$~m/s).

The possibility of contacting many electrodes on the exfoliated Bi$_2$Te$_3$ flakes combined with the reproducibility as observed in the length dependence of the junctions creates an interesting basis for experimental investigations of the still increasing number of theoretical proposals on topological insulator - superconductor devices \cite{Fu2008, Nilsson2008, Tanaka2009}. Interestingly, whereas bulk transport tends to obscure the observation of surface state transport in the normal state, the supercurrent is found to be carried mainly by the surface states. The realization of supercurrents through topological surface states is an important step towards the detection of Majorana fermions.

\textbf{Methods}

By mechanical exfoliation on polycrystalline Bi$_2$Te$_3$ samples, flakes of size up to 50$~\times~$50~\textmu m$^2$ and thicknesses of 20~nm up to 2~\textmu m were transferred to Si substrates. Sputter deposited Nb(200nm)/Pd(5nm) electrodes where defined by lift off techniques using optical photolithography with image reversal photoresist. The Pd layer prevents the Nb from oxidation and the thickness of the Nb being on the order of the flake thickness ensures a superconducting contact ($I_C>30~$mA) between the Nb on the substrate and the flake. The Nb(70~nm) junctions where defined with lift off using e-beam and have a superconducting contact ($I_C>30~$mA) with the Nb electrodes. To obtain transparent contacts, the Bi$_2$Te$_3$ surface is Ar sputter etched \textit{in situ} prior to Nb deposition, while the Bi$_2$Te$_3$ in between the Nb forming the actual junction is unaffected due to the lift off technique. The supercurrent is investigated in shielded cryostats with dedicated home build equipment and the angle dependent magnetoresistance is measured at the High Field Magnet Laboratory (Nijmegen) in a 33~T magnet. 

\textbf{Acknowledgments}

We acknowledge useful discussions with C.W.J. Beenakker, B.C. Kaas, M. Fuhrer, C.G. Molenaar, L. Molenkamp, N. Nagaosa, Y.V. Nazarov and Y. Tanaka. This work is supported by the Netherlands Organization for Scientific Research (NWO) through VIDI and VICI grants, by the Dutch FOM foundation, and by the Australian Research Council through a Discovery project.

\section{- Supplementary Information -}

\textit{Theoretical models for Josephson current}

Any hybrid structure containing superconductors can be described on the basis of the Gor'kov equations \cite{Gorkov}. In practice, these equations are typically simplified by a quasi-classical approximation, which is justified as long as the Fermi-wavelength is much smaller than other length scales in the problem. For superconductor - normal metal - superconductor (SNS) Josephson junctions Eilenberger quasi-classical equations \cite{Eilenberger} are used when the elastic mean free path $l_e$ is larger than the length $L$ and the coherence length $\xi$. The electronic transport in this clean limit is ballistic across the N layer. In the dirty limit of $l_e \ll L, \xi$, transport is diffusive and the Usadel equations \cite{Usadel} are used. When the transparency between the S and N layers is not unity, additional insulating barriers (I) are typically included.

\textit{Eilenberger theory fit}

The clean limit theory on the basis of the Gor'kov equations for short SINIS junctions with arbitrary barrier transparency $D$ \cite{Brinkman2000} was generalized \cite{Bergeret, Galaktionov2002} for arbitrary junction length on the basis of Eilenberger equations. The supercurrent density $J$ is found to be \cite{Galaktionov2002}
\begin{equation}
J=\frac{2}{\pi}ek_F^2 k_B T \textrm{sin}\chi\sum_{\omega_n > 0}\int_0^1 {\mu d\mu \frac{t_1(\mu) t_2(\mu)}{Q^{1/2}(\chi,\mu )}},
\label{J}
\end{equation}
where $\mu=k_x/k_F$, $t_{1,2}=D_{1,2}/(2-D_{1,2})$, and
\begin{equation}
Q=\left[ t_1t_2 \textrm{cos} \chi+ \left( 1+ \left( t_1t_2+1\right)\frac{\omega_n^2}{\Delta^2} \right)\textrm{cosh} \frac{2 \omega_n L}{\mu \hbar v_F} + \left(t_1+t_2 \right)\frac{\omega_n \Omega_n}{\Delta^2}\textrm{sinh}\frac{2 \omega_n L}{\mu \hbar v_F} \right]^2-\left(1-t_1^2\right)\left(1-t_2^2\right)\frac{\Omega_n^4}{\Delta^4},
\end{equation}
where the Matsubara frequency is given by $\omega_n=2\pi k_B T (2n+1)$, and $\Omega_n=\sqrt{\omega_n^2+\Delta^2}$. $\Delta$ is the gap in the S electrodes, $\chi$ the phase difference across the junction, while $v_F$ is the Fermi velocity of the normal metal interlayer. The integral runs over all trajectory directions and can be adjusted to actual junction geometries.

Eq. (\ref{J}) was evaluated as function of junction length and fitted to the measured critical current density. Since the prefactors in Eq. (\ref{J}) implicitly contain the normal state resistance, which is not known for our junctions due to the bulk shunt, we left the overall scale of $J$ free in the fit. Subsequently, the best fit to the data at 1.6 K was obtained for $\xi=\frac{\hbar v_F}{2 \pi k_B T}=75$ nm. It was found numerically that the value of the barrier transparencies in the symmetric case had no influence on the fitting value for $\xi$. 

The temperature dependence of the critical current was calculated using Eq. (\ref{J}) and the obtained coherence length. The fit to the measured data is excellent, considering that only the overall scale of $J$ was free in this case. 

In fact, the overall scaling factor of the critical current in Eq. (1) can be estimated as well. The transparency of the interfaces between the topological insulator and the superconductor are important in this respect. The high transparency of our interfaces can be determined from the $I(V)$ characteristic. The excess current in the $I(V)$ characteristic is about 67$\%$ of the critical current $I_c$. In the Blonder-Tinkham-Klapwijk model \cite{Blonder1982}, this gives a barrier strength of about $Z = 0.6$. For these high transparencies, at the lowest temperatures and for the 50 nm junction, Eq. (1) provides an $I_cR_N$ product of the order of 1-2 mV. However the 3D bulk shunt will strongly reduce this value by decreasing $R_n$. From the Shubnikov de Haas oscillations we can estimate the surface to bulk resistance ratio. The surface resistance can be found through the amplitude of $R_{SDH} = R_c \frac{\lambda}{\sinh{\lambda}} e^{-\lambda _D}$. With $\frac{\lambda}{\sinh{\lambda}} e^{-\lambda _D}$ = 0.77 $\Omega$, we estimate a surface resistance of about 2$\%$ of the total resistance. Thus, the estimated surface resistance is approximately 29 $\Omega$, which together with $I_c$ = 32 \textmu A results in $I_cR_N$ = 1 mV, agreeing with the Eilenberger model. This quantitative agreement between model and measurements underlines the conclusion that supercurrent is flowing through the ballistic channels of the topological surfaces states, shunted by a normal state bulk conduction.

\textit{Usadel theory fit}

The Usadel equation \cite{Usadel} for the S and N layers in a diffusive SNS junction can be written as
\begin{equation}
\Phi_{S,N}=\Delta_{S,N}+\xi^2_{S,N}\frac{\pi k_B T_c}{\omega_n G_{S,N}}\frac{d}{dx}\left( G^2_{S,N}\frac{d}{dx}\Phi_{S,N}\right),
\end{equation}
where $\Phi$ is defined in terms of the normal Green's function $G$ and the anomalous Green's function $F$ by $\Phi G= \omega_n F$. The normalization condition $FF^*+G^2=1$ then gives
\begin{equation}
G_{S,N}=\frac{\omega_n}{\sqrt{\omega_n^2+\Phi_{S,N}\Phi_{S,N}^*}}
\end{equation}
The coherence length is given by $\xi=\sqrt{\frac{\hbar D}{2 \pi k_B T}}$ where $D=v_F l_e/3$ is the diffusion constant.

The pair potentials $\Delta_{S,N}$ are given by
\begin{equation}
\Delta_{S,N} \textrm{ln} \frac{T}{T_{cS,N}}+2\pi k_B T \sum_{\omega_n>0} \frac{\Delta_{S,N}-\Phi_{S,N}G_{S,N}}{\omega_n}=0
\end{equation}

In the dirty limit, Zaitsev's effective boundary conditions for quasi-classical Green's functions were simplified by Kupriyanov and Lukichev \cite{Kupriyanov}. When $\Phi$ and $G$ are found using these boundary conditions, finally the supercurrent density can be obtained from
\begin{equation}
J=\frac{2\pi k_B T}{e \rho_N} \textrm{Im} \sum_{\omega_n>0}\frac{G^2_N}{\omega_n^2} \Phi_N \frac{d}{dx}\Phi_N,
\end{equation}
where $\rho_N$ is the N layer resistivity.

For junctions with arbitrary length and arbitrary barrier transparency, no analytical expressions exist for the Green's functions. Therefore a numerical code was used to fit the data. In the effective boundary conditions \cite{Kupriyanov}, two parameters play a role, $\gamma=\frac{\rho_S \xi_S}{\rho_N \xi_N}$ and $\gamma_B = \frac{2l_e}{3 \xi_N}\left< \frac{1-D}{D} \right>$, where the average of the transparencies takes place over all trajectory angles. For Nb as S electrode and Bi$_2$Te$_3$ as N interlayer, $\gamma \ll 1$ because of the lower resistivity $\rho$ of Nb as compared to Bi$_2$Te$_3$. The junctions transparency is not known a priori, but from the voltage drop over a barrier in the normal state, as well as the large amount of excess current (more than 50\% of the critical current) in the current-voltage characteristics of the superconducting state, a conservative estimate gives $D \gtrsim 0.5$, which implies $\gamma_B \lesssim 1$. Within this parameter range (or even outside the range) no consistent fit could be made to the data. Figure 6a shows the fit to the temperature dependence of the critical current for $\gamma =0.1$, $\gamma_B=1$ and $\xi(T_c)=\sqrt{\frac{\hbar D}{2 \pi k_B T_c}}=21$ nm, the latter value as obtained from fitting the length dependence of the junction.

\end{document}